\documentclass[onecolumn,12pt]{IEEEtran}
\usepackage{mathrsfs}
\usepackage{amsmath}
\usepackage{amsthm}
\usepackage{color}
\usepackage{bbm}
\usepackage{cases}
\usepackage{amsfonts}
\usepackage{amssymb}

\usepackage{booktabs}
\usepackage{graphicx}
\usepackage{cases}
\usepackage{subfigure}
\usepackage[section]{placeins}
\usepackage{caption}
\usepackage{epstopdf}
\usepackage{setspace}
\usepackage{mathtools,amsthm}
\usepackage{algorithm}
\usepackage{algorithmicx}
\usepackage{algpseudocode}
\usepackage{balance}
\usepackage{float}
\usepackage{cite}
\usepackage{subfigure}
\usepackage{array}
\usepackage{picins}
\floatname{algorithm}{Algorithm}

\usepackage{fancyhdr}

\newtheoremstyle{case}{}{}{}{}{}{:}{ }{}
\theoremstyle{case}

\newcommand\old[1]{}

\newtheorem{Proof of Theorem}{Proof of Theorem}
\begin{document}
\title{\huge{Resilient Event-Triggered Control of Vehicle Platoon Under DoS Attacks and Parameter Uncertainty}}
\author{Qiaoni Han, \emph{Member, IEEE}, Jianguo Ma, Zhiqiang Zuo, \emph{Senior Member, IEEE}, Xiaocheng Wang,\\ Bo Yang, \emph{Senior Member, IEEE}, and Xinping Guan, \emph{Fellow, IEEE}
\thanks{This work was supported in part by the National Natural Science Foundation of China under Grant 61803218 and Grant 61973230, in part by the Open Research Programs of the Key Laboratory of System Control and Information Processing, Ministry of Education, under Grant Scip202101, and of the State Key Laboratory of Automotive Simulation and Control under Grant 20210217. \emph{(Corresponding author: Qiaoni Han.)}

Qiaoni Han, Jianguo Ma, and Zhiqiang Zuo are with the Tianjin Key Laboratory of Intelligent Unmanned Swarm Technology and System, School of Electrical
and Information Engineering, Tianjin University, Tianjin 300072, China (e-mail: qnhan@tju.edu.cn; mjg1895@tju.edu.cn; zqzuo@tju.edu.cn).

Qiaoni Han is also with the Key Laboratory of System Control and Information Processing, Ministry of Education, Shanghai 200240, China.

Xiaocheng Wang is with the Tianjin Key Laboratory of Wireless Mobile Communications and Power Transmission, College of Electronic and Communication Engineering Tianjin Normal University, Tianjin 300387, China (e-mail: xcwang@tjnu.edu.cn).

Bo Yang and Xinping Guan are with the Department of Automation, Shanghai Jiao Tong University, Shanghai 200240, China, and also with the Key Laboratory of System Control and Information Processing, Ministry of Education of China, Shanghai 200240, China (e-mail: bo.yang@sjtu.edu.cn; xpguan@sjtu.edu.cn).}}
\markboth{Manuscript}{} \maketitle

\thispagestyle{fancy}
\fancyhead[L]{DOI: 10.1109/TIV.2024.3410314}

\begin{abstract}
This paper investigates the problem of dynamic event-triggered platoon control for intelligent vehicles (IVs) under denial of service (DoS) attacks and parameter uncertainty. DoS attacks disrupt vehicle-to-vehicle (V2V) communications, leading to the destabilization of vehicle formations. To alleviate the burden of the V2V communication network and enhance the tracking performance in the presence of DoS attacks and parameter uncertainty, a resilient and dynamic event-triggered mechanism is proposed. In contrast to the static event-triggering mechanism (STEM), this approach leverages the internal dynamic variable to further save communication resources. Subsequently, a method is developed for designing the desired triggering mechanism. Following this, a co-design framework is constructed to guarantee robust and resilient control against DoS attacks, with the analysis of eliminating Zeno behavior. Lastly, extensive simulations are presented to show the superiority of the proposed method in terms of enhancing platoon resilience and robustness and improving communication efficiency.
\end{abstract}

\begin{spacing}{1.5}
\section{Introduction}
Intelligent vehicles (IVs), which perceive their surrounding environment and perform collaborative behavior through onboard sensors and vehicle-to-vehicle (V2V) communication elements, have been recognized as a revolutionary force in transportation systems\cite{10247090}. The vehicle group consists of intelligent vehicles that have a high degree of autonomy and information exchange on the vehicular ad-hoc network (VANET). They demonstrate a natural cooperation advantage, which can realize high-level traffic modes such as vehicle platoon, and mitigate the escalating congestion and traffic accidents in today's cities. Numerous studies have proven that vehicle platooning can significantly improve road safety, traffic flow, and fuel efficiency\cite{maiti2019impact,103340555,rios2018impact}.

Although data exchange in VANET facilitates improved control performance for vehicular platoons, the context of networked communication gives rise to two critical problems for VANET-based platoon systems: 1) cyber attacks on the information transmission between different vehicles, and 2) the limited bandwidth of VANET.

While V2V communications enable IVs to gather the state information of their neighboring vehicles, the implementation of a wireless network also exposes vehicle platoon systems vulnerable to external malicious attacks\cite{amoozadeh2015security}. Typical attacks considered in secure platoon control encompass denial of service (DoS) attacks and deception attacks. DoS attacks on IVs are usually carried out by disrupting the radio frequency or flooding the V2V communication access with excessive requests, aiming to overwhelm the communication resources and impede legitimate vehicles from connecting\cite{666888}. In contrast, deception attacks typically compromise data trustworthiness or integrity. Typical deception strategies encompass replay attacks and false data injection attacks\cite{fu2021data,xu2020robust}.

In addressing the cyber security challenge, recent research has delved into secure platoon control methods\cite{ju2022survey}. The existing attack mitigation and elimination methods for safeguarding vehicle platoons are generally classified into three types: prevention-based methods, detection-based methods, and resilience-based methods. Among these, prevention-based methods usually rely on information assurance techniques, including data authentication and cryptographic algorithms\cite{10099187}. Detection-based methods are designed to identify the presence of specific types of attacks\cite{9908023}. From the perspective of system control and estimation, the emphasis is to construct an appropriate state observer or estimator on each vehicle, based on either Kalman filtering theory, neural networks, or machine learning techniques, to estimate the dynamic state of the vehicle in real-time and trigger a warning upon detecting attacks.

Resilience-based approaches generally depend on pre-designed controllers that are resilient to the impact of attacks\cite{94993}. In this context, resilience characterizes the capability to remain in operation in IVs under malicious attacks. Evidently, prevention-based methods and detection-based methods are foundational in establishing protection measures and security monitoring, whereas resilience-based approaches intend to preserve the security requirements of vehicle platoon systems by achieving some control objectives. In turn, the analysis results of control system resilience also have a certain guiding significance for the other two methods. In\cite{xu2022resilient}, a distributed security controller is proposed to relieve the influence of DoS attacks, and the tracking performance for connected vehicles based on sampled data is guaranteed by the switching delay system method.\cite{ge2022resilient} designed a resilient control law to ensure platoon scalability, individual vehicle stability, and attack resilience. In\cite{zhao2021resilient}, a resilient control framework was developed to ensure the disturbance attenuation performance of vehicular cyber-physical systems under DoS attacks. Moreover, an adaptive synchronization cooperative control method was developed in\cite{petrillo2020secure} to withstand various types of attacks. While the above secure platoon control algorithms achieve resilience against DoS attacks, they ignore the issue of limited communication resources in VANET.

The carrying capacity of VANET bandwidth is a primary concern that constrains the quantity and quality of data transmission. To address the issue of limited communication resources, the event-triggered mechanism (ETM) has emerged to reduce the frequency of data transmissions in networked control systems and multi-agent systems\cite{chen2022command,zhang2022resilient}. In the realm of event-triggered platooning control, the majority of studies focus on static triggering conditions, where the triggering parameters are pre-designed and remain constant, such as those in\cite{xiao2022resource,1013064999}, or can be adjusted according to the communication bandwidth within the selectable range of static parameters, such as those in \cite{ge2021dynamic,xiao2021dynamic}. Recently, the dynamic event-triggered mechanism (DETM) introduced in \cite{girard2014dynamic} has been expanded to multi-agent systems to provide a more adaptable and flexible sampling schedule\cite{10302179}. In \cite{8468237}, a distributed control protocol leveraging a DETM was formulated to address the leaderless consensus problem. Furthermore, \cite{8485753} introduced a dynamic event-based control law to tackle the average consensus issue across undirected graphs. The problem of applying DETM to average consensus or leaderless consensus control could not be applied to vehicle platoon control in cyber attack scenarios.

In addition to the aforementioned critical problems, another aspect that has received less attention in platoon control design is the robustness of the platoon under parameter uncertainty. In practical platoon control, parameter uncertainty inevitably exists due to neglected high-order dynamics and environmental disturbances. To mitigate prediction uncertainty,\cite{feng2021robust} proposed a robust platoon control framework to mitigate prediction uncertainty by dynamic feedforward control and feedback control. For platoon control under parameter uncertainty and communication delay, a distributed robust proportional-integral-derivative controller was introduced in \cite{fiengo2019distributed} to enhance the platoon robustness stability under parameter uncertainty and communication delay. Up to now, there have been few results on the application of DETM to vehicle platoon control, especially concerning robust control in the presence of parameter uncertainty.

In this paper, our investigation focuses on the application of DTEM for vehicle platoon control systems under DoS attacks and parameter uncertainty. The contributions of this paper can be summarized as follows:
\begin{itemize}
\item[$\bullet$]Considering parameter uncertainty and DoS attacks, the problem of applying DETM to vehicle platoon systems with directed communication topology is studied.
\item[$\bullet$]Different from the SETM for the vehicle platoon system, a resilient and dynamic ETM with an internal dynamic variable that can be adjusted based on the state estimation error and the neighborhood tracking error to save communication resources and withstand the impact of DoS attacks.
\item[$\bullet$]A co-design framework of a robust controller and a DETM is suggested. Within this framework, the robust controller is designed to ensure robustness to parameter uncertainty. Furthermore, the relevant parameters of the DETM are obtained in the design process and the resilience of DoS attacks is theoretically analyzed and verified by simulations.
\end{itemize}

The rest of this paper is structured as follows. Section II describes the model of parameter uncertainty and DoS attacks, along with the design of a resilient and dynamic ETM. Section III constructs a co-design framework of a robust controller and a DETM while excluding the Zeno phenomenon. Section IV gives simulation results, demonstrating that the proposed design method effectively maintains the performance of control in the presence of DoS attacks and parameter uncertainty while saving communication resources. Section V summarizes the findings of this paper.

\textit{Notations:} The space of $n\times m$ real matrices and the set of natural numbers are denoted as $\mathbb{R} ^{n\times m}$ and $\mathbb{N}$, respectively. For a matrix $\textit{X}\in\mathbb{R}^{n\times n}$, $\textit{X}>0$ means that $\textit{X}$ is positive definite. The transpose of matrix $\textit{X}$ is denoted by $\textit{X}^T$ and $Tr(\textit{X})=\textit{X}+\textit{X}^T$.
\section{PROBLEM FORMULATION}
\subsection{Longitudinal Vehicle Platoon Model}
Consider a connected vehicle system consisting of one leader vehicle and $N$ following vehicles. Let $p_{i}( t )$ , $v_{i}\left ( t \right )$ and $a_{i}(t)$ denote the longitudinal position, velocity, and acceleration of the following vehicle $i$. According to\cite{rajamani2011vehicle}, through the utilization of nonlinear compensation, the first-order inertial transfer function can approximate the longitudinal vehicle dynamics, thereby allowing it to be described as:
 \begin{eqnarray}
&&\hspace{-0.9cm}\dot{p}_i(t) =v_i(t),\nonumber\\
&&\hspace{-0.9cm}\dot{v}_i(t) =a_i(t),\nonumber\\
&&\hspace{-0.9cm}\dot{a}_i(t) =-\frac{1}{\tau } a_i(t)+\frac{1}{\tau }u_i(t).\nonumber
\end{eqnarray}
Note that the value of the power-train time constant $\tau$ depends on various driving conditions \cite{li2017robustness}, the bounded parameter uncertainty on $\tau$ is employed, i.e.,
 \begin{eqnarray}
&&\hspace{-1.5cm}{\tau } =\bar{\tau } +\Delta \tau,\nonumber
\end{eqnarray}
where $\bar{\tau }$ is the nominal value, and $\Delta \tau $ is the relevant uncertainty. For convenience we set
 \begin{eqnarray}
&&\hspace{-0.9cm}  \frac{1}{\tau}= \frac{1}{\bar{\tau}+\Delta \tau }=\bar{\varpi }+\Delta\varpi> 0.\label{20}
\end{eqnarray}
Subsequently, the vehicle model with bounded parameter uncertainty can be expressed as
\begin{align}
\hspace{-0.9cm}\dot{x}_i(t) &=\tilde{A}x_i(t)+\tilde{B}u_i(t)\nonumber\\
                              &=(\bar{A}+\Delta A)x_i(t)+(\bar{B}+\Delta B)u_i(t),\nonumber
\end{align}
where $x_{i}(t)=\left [ p_{i}(t),v_{i}(t),a_{i}(t)  \right ]^{T}\in \mathbb{R} ^{3\times 1}$ and
\begin{eqnarray*}
&&\hspace{-0.9cm}\bar{A}=\begin{bmatrix} 0 & 1 & 0\\ 0 & 0 & 1\\0  &  0&-\bar{\varpi }  \end{bmatrix},\Delta A=\begin{bmatrix}  0& 0 &0 \\ 0 &  0&0 \\ 0 &0  &-\Delta \varpi \end{bmatrix},
\end{eqnarray*}
\begin{eqnarray*}
&&\hspace{-3.5cm}\bar{B}=\begin{bmatrix} 0\\0\\\bar{\varpi } \end{bmatrix},\Delta B=\begin{bmatrix} 0\\0\\\Delta \varpi  \end{bmatrix}.
\end{eqnarray*}
Let $p_{0}\left ( t \right )$ and $v_{0}\left ( t \right )$ denote the longitudinal position and velocity of the leader vehicle 0.
Then, the longitudinal dynamics of the leader vehicle can be described by
\begin{eqnarray}
&&\hspace{-0.9cm}\dot{x}_0(t) =\tilde{A}x_0(t),\nonumber
\end{eqnarray}
where $x_{0}(t)=\left [ p_{0}(t),v_{0}(t),a_{0}(t)  \right ]^{T}\in \mathbb{R} ^{3\times 1}$.

The fixed spacing distance policy is adopted in this work. Let $l _{i,j}=[(i-j)*l  ,0,0] ^{T} \in \mathbb{R} ^{3\times 1}$, where $l$ represents the desired inter-vehicle spacing with $(i-j)*l$ denoting the prescribed longitudinal distance between vehicles $i$ and $j$.
\subsection{V2V Communication}
The communication topology of $N$ follower vehicles can be modeled by a directed digraph $\mathcal{G}=(\mathcal{V}, \mathcal{E},\Pi )$, where $\mathcal{V} = \{ 1,2,\cdots,N \}$ represents a set of $N$ nodes, $\mathcal{E}\subseteq \mathcal{V} \times \mathcal{V}$ stands for a set of edges and $(i,j)$ indicates directional edge from node $i$ to node $j$. $\Pi =[a_{ij}]\in\mathbb{R} ^{N\times N}$ is the adjacency matrix in which $a_{ij}=1$ if $(\mathcal{V}_i,\mathcal{V}_j)\in\mathcal{E}$, and $a_{ij}=0$, otherwise. The Laplacian matrix of the digraph $\mathcal{G}$ is defined as $\mathcal{L}=\mathcal{D}-\Pi $, where $\mathcal{D}=$ diag$\{d_1,d_2,\cdots,d_N\}$ with $d_i=\sum_{j=1}^{N} a_{ij}$. The pinning matrix is defined as $H$ = diag$\{h_1,h_2,\cdots,h_N\}$, in which $h_i=1$ if follower $i$ can receive information from leader 0, and $h_i = 0$, otherwise. Define $\tilde{\mathcal{G} }$ be the communication graph that includes a leader node and $N$ follower nodes. Correspondingly, $\mathcal{H}=\mathcal{L}+H$ represents an information flow matrix that delineates the algebraic characteristics of $\tilde{\mathcal{G} }$.

\emph{Assumption 3: }\label{a1} There is a spanning tree rooted at the leader node 0 in the communication digraph $\tilde{\mathcal{G} }$.
\subsection{Model of DoS Attacks}
\begin{figure}[h]
\includegraphics[scale=0.35]{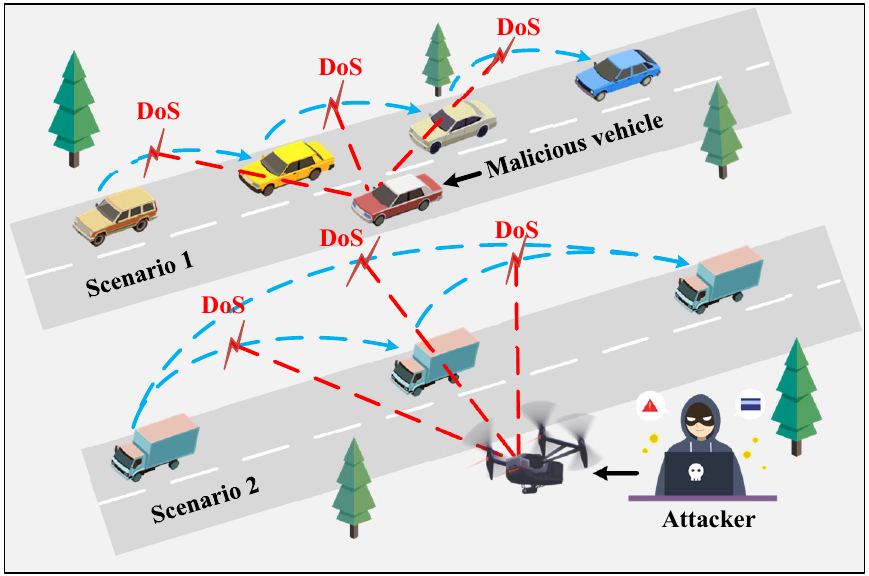}\centering
\caption{Scenarios of DoS attacks on vehicular platoon.}
\label{figure}
\end{figure}
As shown in Fig. 1, a typical attack scenario is when a malicious vehicle drives to the side of the platoon to prevent the required exchange of information between vehicles deliberately. Another potential attack scenario involves a signal jammer installed on a drone hovering over the platoon. The attacker can interfere with V2V communication channels whenever a vehicle transmits information. These malicious attacks lead to intermittent disruption of real-time V2V communication, resulting in the loss of vehicle packets on the network. This paper focuses on examining the impact of such malicious DoS attacks.

The c-th attacked interval is described as
\begin{eqnarray}
&&\hspace{-0.9cm}A_{c} =a_{c} \bigcup \left [ a_{c}, a_{c}+d_c \right ),c\in \mathbb{\mathbb{N} },\nonumber
\end{eqnarray}
where $a_{c}$ and $a_{c} + d_c$ represent the begining and end instants of attack. During $\left [ t_{1}, t_{2}\right )$,  The entire active interval of attack is
\begin{eqnarray}
&&\hspace{-0.9cm}\mathcal{A}\left (t_{1},t_{2} \right ) =\left \{ \bigcup_{c\in\mathbb{N}  } A_{c}  \right \} \bigcap \left [ t_{1}, t_{2}\right ).\nonumber
\end{eqnarray}
Obviously, the entire secure communication interval during $\left [ t_{1}, t_{2}\right )$ is
\begin{eqnarray}
&&\hspace{-0.9cm}\mathcal{S}\left (t_{1},t_{2}  \right ) =  \left [ t_{1} , t_{2}\right )\backslash\mathcal{A}\left (t_{1},t_{2} \right ).\nonumber
\end{eqnarray}
Furthermore, let $\mathcal{N} \left ( t_{1},t_{2} \right )$  be the total number of DoS off-to-on occurring in interval $\left [ t_{1}, t_{2}\right )$.

\emph{Assumption 2: }\cite{de2015input}\label{a1} There exist four scalars $T_{1}>0$, $T_{2}>0$, $D_{1}>0$ and $D_{2}>1$, such that for $t_{2}\ge t_{1}\ge0$,
\begin{eqnarray}
&&\hspace{-0.9cm}\left |\mathcal{A} \left ( t_{1}, t_{2} \right )  \right |  \le D_{1}+\frac{t_{2}-t_{1}}{D_{2}},\nonumber\\
&&\hspace{-0.9cm}\mathcal{N} \left ( t_{1},t_{2} \right )\le T_{1}+\frac{t_{2}-t_{1}}{T_{2}}.\nonumber
\end{eqnarray}

\subsection{Event-Based Distributed Platooning Control Law}
To save communication resources, we construct the following event-based distributed platooning control law:
\begin{eqnarray}
&&\hspace{-1.3cm}u_{i}( t )=K\big[\sum_{j=1}^{N}  a_{ij}(\hat{x}_{i}(t)-\hat{x}_{j}(t)-l _{i,j} )\nonumber\\
&&\hspace{0.8cm}+ h_{i}(\hat{x}_{i}(t)-\hat{x}_{0}(t)-l _{i,0})  \big],\label{1}
\end{eqnarray}
where
\begin{eqnarray}
&&\hspace{-0.5cm}\hat{x} _{i} (t)=x_{i} (t),t\in \left \{ t_{k}^{i}  \right \}, \nonumber\\
&&\hspace{-0.5cm}\dot{\hat{x}} _{i} (t)=\bar{A}\hat{x}_{i} (t),t\in \left [ t_{k}^{i},t_{k+1}^{i} \right ) ,k\in \mathbb{N}.\label{2}
\end{eqnarray}
$K\in \mathbb{R} ^{3\times 1}$ is the controller gain to be designed and $t_{k}^{i}$ denotes the k-th triggering sample instant of vehicle $i$ with $t_{0}^{i}$ = 0. (\ref{2}) is an open-loop state estimator of state $x_i(t)$ built on vehicle $i$ and its neighboring vehicles to generate the state estimate $\hat{x} _{i} (t)$.
\subsection{Resilient and Dynamic Event-Triggered Mechanism}
\begin{figure}[H]
\includegraphics[scale=0.15]{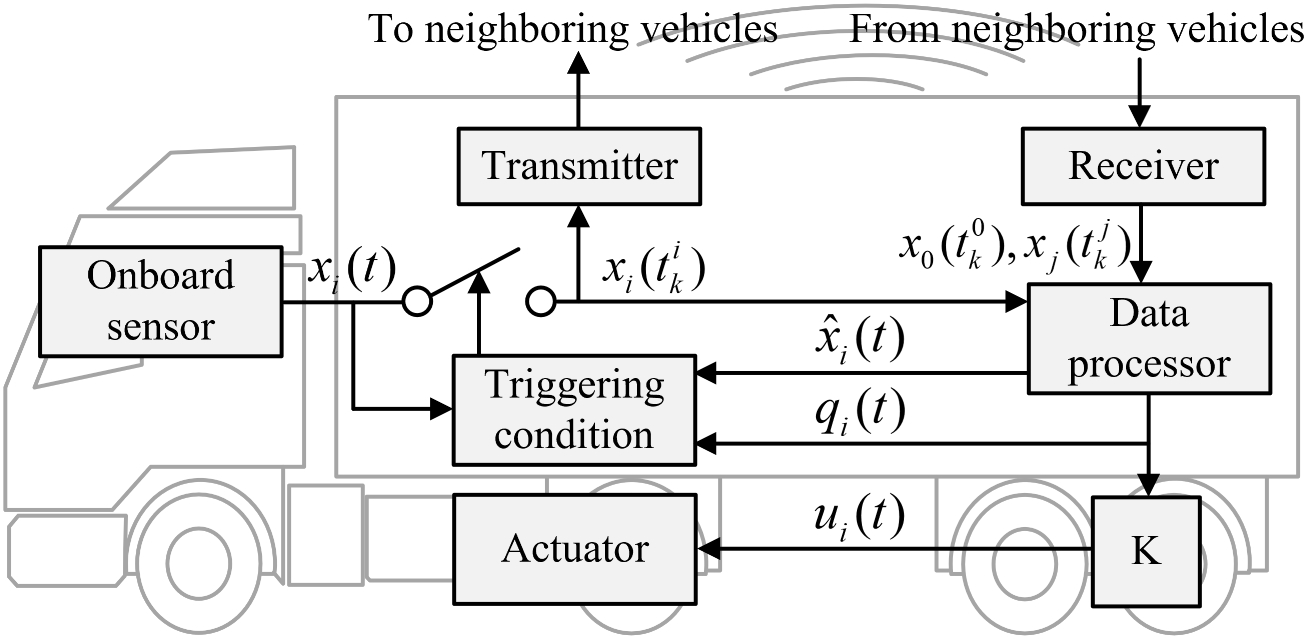}\centering
\caption{Longitudinal event-triggered control mechanism.}
\end{figure}
In this section, we design a resilient and dynamic ETM as depicted in Fig. 2 to accomplish the desired distributed platooning control objective while under DoS attacks.

For vehicle $i$, we define the state estimation error as:
\begin{eqnarray}
&&\hspace{-0.9cm}\varepsilon _{i}(t)= \hat{x} _{i} (t)-x _{i} (t),\label{5}
\end{eqnarray}
and the neighborhood tracking error as:
\begin{eqnarray}
&&\hspace{-0.9cm}q_{i}(t)=\sum_{j=1}^{N}  a_{ij}(\hat{x}_{i}(t)-\hat{x}_{j}(t)-l_{i,j} )\nonumber\\
&&\hspace{0.8cm}+ h_{i}(\hat{x}_{i}(t)-\hat{x}_{0}(t)-l_{i,0}).\nonumber
\end{eqnarray}
Then, the triggering instance $t_{k+1}^{i}$ is given based on the following triggering law:
\begin{eqnarray}
&&\hspace{-0.9cm}t_{k+1}^{i} = \mathrm{inf} \left \{ t> t_{k}^{i}|\pi _{i} \left ( t \right ) \ge 0 \right \},\label{3}
\end{eqnarray}
where $\pi _{i} \left ( t \right )$ indicates the dynamic triggering condition and satisfies
\begin{eqnarray}
&&\hspace{-0.9cm}\pi _{i} \left ( t \right ) =\beta _{1} \left \| \varepsilon _{i}(t)  \right \|^{2}  -\beta _{2} \left \| q _{i}(t)  \right \|^{2}-\varphi _{i} \theta _{i} (t),\label{4}
\end{eqnarray}
where $\theta _{i} (t)$ is the internal dynamic variable that satisfies the following differential equation:
\begin{eqnarray}
&&\hspace{-0.9cm}\dot{\theta} _{i} (t)=-\varrho _{i} \theta_{i} (t)-\eta _{i} \left ( \beta _{1} \left \| \varepsilon _{i}(t)  \right \|^{2}  -\beta _{2} \left \| q _{i}(t)  \right \|^{2} \right ),\label{12}
\end{eqnarray}
where $\varphi _{i} \ge 0,\eta _{i} \ge 0,1> \varrho_{i} > \varphi _{i}(1-\eta _{i}) \ge 0$. $\beta _{1}$ and $\beta _{2}$ will be given later. If the triggering law $(\ref{3})$ is satisfied, the state information of vehicle $i$ is transmitted through the network to its neighboring vehicles. Note that the internal dynamic variable $\theta _{i} (t)$ is a crucial element of the dynamic triggering condition (\ref{4}), which is inspired from\cite{girard2014dynamic}. As the parameter $\theta _{i} (t)$ approaches zero, the dynamic triggering condition degenerates to a static triggering one:
\begin{eqnarray}
&&\hspace{-0.9cm}\pi _{i} \left ( t \right ) =\beta _{1} \left \| \varepsilon _{i}(t)  \right \|^{2}  -\beta _{2} \left \| q _{i}(t)  \right \|^{2}.\label{21}
\end{eqnarray}
Considering the impact of DoS attacks, the DETM keeps ineffective triggering and may cause the Zeno phenomenon. To this end, a resilient and dynamic ETM is designed as:
\begin{eqnarray}
&&\hspace{-0.9cm}t_{k+1}^{i}=\left\{\begin{array}{cccc}
&&\hspace{-0.9cm}t_{k}^{i} \, \mathrm{satisfying } \,(\ref{3}),&\mathrm{if} \,t_{k}^{i}\in \mathcal{S} \left ( t_{1}, t_{2} \right )\\
&&\hspace{-2.43cm}\ t_{k}^{i}+h,&\,\mathrm{if} \,t_{k}^{i}\in \mathcal{A} \left ( t_{1}, t_{2} \right ) \label{11}
\end{array} \right.
\end{eqnarray}
As shown in Fig. 3, during the secure communication interval $\mathcal{S} \left ( t_{1}, t_{2} \right )$, vehicle $i$ samples its state information through dynamic event-triggering scheme, while in the active interval of attack $\mathcal{A} \left ( t_{1}, t_{2} \right )$, vehicle $i$ takes $h$ as the sampling period till communication is restored. $\sigma_{c}$ is the prolonged affected interval of c-th DoS attack and $\sigma_{c}\le h$ due to periodic sampling mechanism.
During $[t_{1},t_{2})$, the entire prolonged affected interval can be expressed as:
\begin{eqnarray}
&&\hspace{-0.9cm}\sigma (t_{1},t_{2} )=\bigcup_{c\in \mathbb{N} }\sigma_{c}.\nonumber
\end{eqnarray}
Then, the entire affected interval of attacks becomes
\begin{eqnarray}
&&\hspace{-0.9cm}\mathcal{A} ^{*} \left ( t_{1} ,t_{2} \right ) =\mathcal{A}  \left ( t_{1} ,t_{2} \right ) \bigcup \sigma (t_{1},t_{2} ).\nonumber
\end{eqnarray}
Apparently, the entire unaffected interval of attacks is
\begin{eqnarray}
&&\hspace{-0.9cm}\mathcal{S}^{*}\left (t_{1},t_{2}  \right ) =  \left [ t_{1} , t_{2}\right )\backslash\mathcal{A}^{*}\left (t_{1},t_{2} \right ).\nonumber
\end{eqnarray}
\begin{figure}[H]
\includegraphics[scale=0.14]{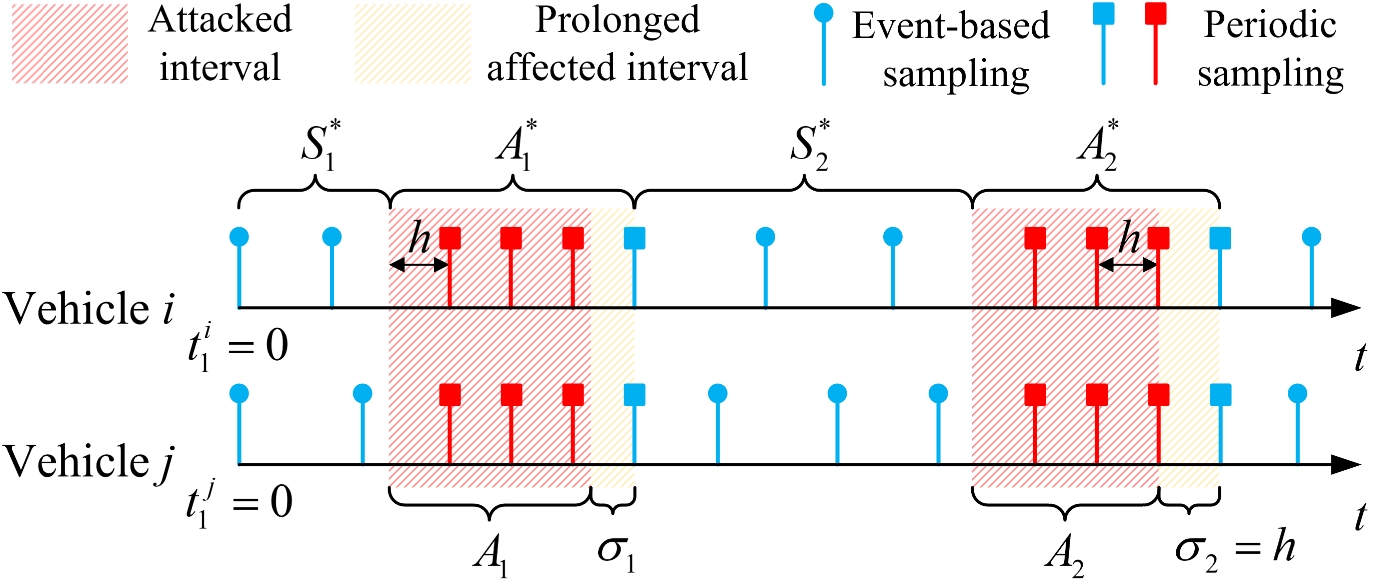}\centering
\caption{Relationship between attacked intervals and affected intervals.}
\end{figure}
\subsection{Tracking Error Dynamics}
Define the tracking error of each following vehicle $i$ be
\begin{eqnarray}
&&\hspace{-0.9cm}e_{i}(t)= x_{i}(t)-x_{0}(t)-l _{i,0}.\label{6}
\end{eqnarray}
Then, we have
\begin{align}
\hspace{-0.1cm}\dot{e} _{i}(t)&=\tilde{A}e _{i}(t)+\tilde{B}K\big[\sum_{j=1}^{N} a_{ij}(\hat{x}_{i}(t)-\hat{x}_{j}(t)-l _{i,j} )\nonumber\\
                              &\ \ \ + h_{i}(\hat{x}_{i}(t)-\hat{x}_{0}(t)-l _{i,0})  \big].\label{7}
\end{align}
Substituting (\ref{5}) and (\ref{6}) into (\ref{7}), it can be rearranged as a compact form
\begin{eqnarray}
&&\hspace{-1.2cm}\dot{e} (t)=(I_{N} \otimes \tilde{A}-\mathcal{H} \otimes \tilde{B}K)e (t)-(\mathcal{H} \otimes \tilde{B}K)\varepsilon  (t).\label{9}
\end{eqnarray}
where $e(t)= [ e_{1} (t)^{T},e_{2} (t)^{T},\cdots ,e_{N} (t)^{T}  ]^{T} \in \mathbb{R} ^{3N\times 1}$, $\varepsilon (t)=[ \varepsilon _{1} (t)^{T},\varepsilon _{2} (t)^{T},\cdots , \varepsilon _{N} (t)^{T} ] ^{T}\in \mathbb{R} ^{3N\times 1}  $.

During the affected time interval of attack, the input signal required for control updates is interrupted. It is reasonable to set $u_{i}(t)=0 $ and $\dot{\theta} _{i} (t)=0$, $t\in \mathcal{A} ^{*} \left ( t_{1} ,t_{2} \right )$ to maintain the subsequent functions of the DETM. Considering the unaffected time interval $\mathcal{S}^{*}\left (t_{1},t_{2}  \right )$ and the  affected time interval $\mathcal{A} ^{*} \left ( t_{1} ,t_{2} \right )$, system $(\ref{9})$ is re-formulated as
\begin{eqnarray}
&&\hspace{-0.9cm}\dot{e}(t)=\left\{\begin{array}{cccc}
&&\hspace{-1.6cm}(I_{N} \otimes \tilde{A})e(t)-(\mathcal{H}\otimes \tilde{B}K)e (t)\\
&&\hspace{-0.9cm}\,\,\,\,\,\,\,\,-(\mathcal{H}\otimes \tilde{B}K)\varepsilon(t),\,\,t\in\mathcal{S}^{*}\left (t_{1},t_{2}  \right )\\
&&\hspace{-0.9cm}(I_{N} \otimes \tilde{A})e(t),\,\,\,\,\,\,\,\,\,\,\,\,\,\,\,\,\,\,\,t\in\mathcal{A}^{*}\left (t_{1},t_{2}  \right )\label{10}\\
\end{array} \right.
\end{eqnarray}
\section{MAIN RESULTS}
Initially, we give sufficient conditions to ensure the stability of the tracking error system in (\ref{10}) in the absence of parameter uncertainty and provide a congruent transformation method to get the controller gain. Then we establish a co-design framework of a robust controller and a DTEM.
\subsection{Stability Analysis in the Absence of Parameter Uncertainty}
\emph{Theorem 1: }For the vehicle platoon over V2V communication topology $\tilde{\mathcal{G} }$ satisfying Assumption 1, under the platooning control law in $(\ref{1})$ and assuming the absence of parameter uncertainty, for given parameters $\varphi _{i} \ge 0,\eta _{i} \ge 0,1> \varrho_{i} > \varphi _{i}(1-\eta _{i}) \ge 0,$ and $c\ge 0,$ the tracking error system in (\ref{10}) is stable if there exists a positive definite matrice $P\in \mathbb{R} ^{3\times 3}$ and positive scalars $\beta _{1},$ and $\beta _{2}$ satisfying the following inequality,
\begin{align}
\bar\Psi&=\begin{bmatrix} \Xi_{1,1} &-\mathcal{H} \otimes P\bar{B}K &\mathcal{H}^T\otimes I_3\\ -\mathcal{H}^T \otimes K^T\bar{B}^TP & -\frac{\beta_1}{\beta_2}I_{3N}  &\mathcal{H}^T\otimes I_3 \\ \mathcal{H}\otimes I_3 & \mathcal{H}\otimes I_3 &-I_{3N}\end{bmatrix}< 0,\nonumber\\
\label{13}
\end{align}
where $\Xi_{1,1} =Tr(I_N\otimes P\bar{A} -\mathcal{H} \otimes P\bar{B}K)+cI_N\otimes P$, and the parameters of DoS attacks in Assumption 2 satisfy
\begin{eqnarray}
&&\hspace{-0.9cm}T_{2} > \frac{2\ln{\phi  +\left ( \varsigma _{1} +\varsigma _{2}  \right )h } }{\varsigma _{*}}, \nonumber\\
&&\hspace{-0.9cm}D_{2} >\frac{\varsigma _{1}+\varsigma _{2} }{\varsigma _{1}-\varsigma _{*}},\label{16}
\end{eqnarray}
where $\varsigma _{1} =$ min $(c/\lambda _{max} (P),\varrho _i ),0< \varsigma _{*} \le\varsigma _{1},\bar{A}^{T} Q+Q\bar{A}\le \varsigma _{2}I_{n},\phi  $ is the gain scheduler such that $ P\le \phi  Q,Q \le\phi  P$.

\emph{Proof: }
The piecewise Lyapunov functional $W(t)$ is
\begin{eqnarray}
&&\hspace{-0.9cm}W(t)=\left\{\begin{array}{cccc}
&&\hspace{-0.9cm}V_{1} \left ( t \right ) +\sum_{i=1}^{N} \theta _{i} (t),\,\,\,\,\,\,\,t\in \mathcal{S} ^{*} \left ( t_{1} ,t_{2} \right )\nonumber\\
&&\hspace{-0.9cm}V_{2} \left ( t \right ) +\sum_{i=1}^{N} \theta _{i} (t),\,\,\,\,\,\,\,t\in \mathcal{A} ^{*} \left ( t_{1} ,t_{2} \right )\nonumber\\
\end{array} \right.
\end{eqnarray}
For $t\in \mathcal{S} ^{*} ( t_{1} ,t_{2} )$, where $V_{1} ( t  ) =e^{T}  (t) (I_{N}\otimes P )e  (t)$. The time derivative of $W(t)$ is computed as follows:
\begin{align}
\hspace{0cm}\dot W(t)&=2e(t)^T(I_N\otimes P)\dot e(t)+\sum_{i=1}^{N} \dot{\theta} _i(t)\nonumber\\
	                     &=e(t)^T[Tr(I_N\otimes P\bar{A}-\mathcal{H}\otimes P\bar{B}K)]e(t)\nonumber\\
                         &\ \ \ +\sum_{i=1}^{N} \left(-\varrho _{i} \theta_{i} (t)-\eta _{i}  ( \beta _{1} \left \| \varepsilon _{i}(t)  \right \|^{2}  -\beta _{2} \left \| q _{i}(t)  \right \|^{2}  )\right)\nonumber\\
                         &\ \ \ -2e^T(t)[\mathcal{H}\otimes P\bar{B}K]\varepsilon (t)\nonumber\\
                         &=e(t)^T[ Tr(I_N\otimes P\bar{A}-\mathcal{H}\otimes P\bar{B}K)]e(t)\nonumber\\
                         &\ \ \ +\left \| \mathcal{H}\otimes I_3(e(t)+\varepsilon (t))  \right \| ^2-\frac{\beta_1}{\beta_2} \varepsilon (t)^2-\sum_{i=1}^{N} \varrho _i\theta _i(t)\nonumber\\
                         &\ \ \ -2e^T(t)[\mathcal{H}\otimes P\bar{B}K]\varepsilon (t).\nonumber
\end{align}
According to (\ref{13}), it can be further written as:
\begin{align}
\hspace{-0.1cm}\dot{W}(t)&\le-ce^T(t)(I_N\otimes P)e(t)-\sum_{i=1}^{N} \varrho _i\theta _i(t)\nonumber\\
                         &\le-\varsigma _{1}W(t),\label{14}
\end{align}
where $\varsigma _{1} =$ min $(c/\lambda _{max} (P),\varrho _i )$.

For $t\in \mathcal{A} ^{*} \left ( t_{1} ,t_{2} \right )$, where $V_{2} \left ( t \right ) =e^{T} \left ( t \right ) (I_{N}\otimes Q )e \left ( t \right )$, taking the derivative of $W\left ( t \right )$ along the trajectory of the tracking error system $(\ref{10})$, we have
\begin{align}
\hspace{0cm}\dot{W}(t)&=e^T(t)(I_N\otimes (\bar{A}^TQ+Q\bar{A})e(t)\nonumber\\
&\le\varsigma _{2}W(t).\label{15}
\end{align}
Combining $(\ref{14})$ and $(\ref{15})$, after iteration we obtain
\begin{align}
\hspace{0cm}{W}(t)&\le e^{-\varsigma _1(t-a_{c}-d_{c}-h  )} (V_{1}(a_{c}+d_{c}+h)\nonumber\\
                   &\ \ \ +\sum_{i=1}^{N}\theta _{i}(a_{c}+d_{c}+h)  )\nonumber\\
                   &\le\phi e^{-\varsigma _1(t-a_{c}-d_{c}-h  )} e^{d_{c}+h} (V_{2}(a_{c})+\sum_{i=1}^{N}\theta _{i}(a_{c}))\nonumber\\
                   &\le\dots\nonumber\\
                   &\le \phi ^{2\mathcal{N}(0,t) }e^{-\varsigma _1| \mathcal{S}^*(0,t) | }  e^{\varsigma _2| \mathcal{A}^*(0,t) | }W(0),\nonumber
\end{align}
where $|  \mathcal{S} ^*(0,t) |=t- |  \mathcal{A} ^*(0,t) | $ and $|\mathcal{A}^*(0,t)|\le D_1+t/D_2 +(1+\mathcal{N}(0,t) )h$. Finally we get
\begin{align}
\hspace{0cm}{W}(t)&\le e^{2\mathcal{N}(0,t)\ln{\phi  }  } e^{-\varsigma _1t  }e ^{(\varsigma _1 +\varsigma _2 )|  \mathcal{A^*}(0,t) |  }W(0)\nonumber\\
                   &\le e^{2\ln{\phi  +(\varsigma _1+\varsigma _2)(T_1+D_1+1)} }e^{(-\varsigma _1+(\varsigma _1+\varsigma _2)/D_2+\varsigma _*)t}W(0),\nonumber
\end{align}
where $\varsigma ^*=(h+2\ln{\phi } )/T_2$. Under condition $(\ref{16})$ , $-\varsigma _1+(\varsigma _1+\varsigma _2)/D_2+\varsigma _*<0$. Thus, tracking error system in (\ref{10}) is stable under DoS attacks.

\emph{Remark 1: }Within Theorem 1, the controller gain $K$ can be solved using congruent transformation. Denote $P^{-1}=M$, and $KP^{-1}=U$, if there exist positive definite matrices $M\in \mathbb{R} ^{3\times 3}$, and $U\in \mathbb{R} ^{3\times 3}$ satisfying the following inequality,
\begin{align}
\bar\Psi&=\begin{bmatrix} \Xi_{C} &-\mathcal{H} \otimes \bar{B}U &\mathcal{H}^T\otimes M\\ -\mathcal{H}^T \otimes U^T\bar{B}^T & -\frac{\beta_1}{\beta_2}MI_{3N}M  &\mathcal{H}^T\otimes M \\ \mathcal{H}\otimes M & \mathcal{H}\otimes M &-I_{3N}\end{bmatrix}< 0,\nonumber
\end{align}
where $\Xi_{C} =Tr(I_N\otimes \bar{A}M -\mathcal{H} \otimes \bar{B}U)+cI_N\otimes M$. The controller gain $K$ is given by $K=UM^{-1}$.
\subsection{Robust Stability Analysis}
\emph{Lemma 1\cite{briat2014linear}: }Given matrices $F,D$ and $V(t)$ with appropriate dimension, where $V(t)$ is time-varying, for any $\vartheta_i>0$ with $V(t)^TV(t)\le I$, we have
\begin{align}
\hspace{0cm}Tr(FV(t)D)&\le F\Lambda F^T+D\Lambda^{-1} D^T\nonumber\\
                      &\le D\Lambda^{-1} D^T+F\Lambda F^T,\nonumber
\end{align}
with $\Lambda=$ diag$\{\vartheta_1I,\vartheta_2I,\cdots,\vartheta_iI\}$.

\emph{Theorem 2: }For the vehicle platoon over V2V communication topology $\tilde{\mathcal{G} }$ satisfying Assumption 1, under the  platooning control law in $(\ref{1})$. For given constant matrices $F_a,F_b,D_a\in \mathbb{R} ^{3\times 3}$, and $D_b\in \mathbb{R} ^{3\times 1}$, and assuming that the hypotheses of Theorem 1 hold, the tracking error system in (\ref{10}) has robust stability if the following inequality is satisfied:
\begin{align}
\tilde\Psi&=\begin{bmatrix} \Xi_{1,1}+\Xi_A+\Xi_B &\Xi_{1,2} &\mathcal{H}^T\otimes I_3\\ \Xi_{2,1} & -\frac{\beta_1}{\beta_2}I_{3N}  &\mathcal{H}^T\otimes I_3 \\ \mathcal{H}\otimes I_3 & \mathcal{H}\otimes I_3 &-I_{3N}\end{bmatrix}< 0,\label{19}
\end{align}
where
\begin{align}
&\Xi_A=I_N\otimes PF_a(I_N\otimes PF_a)^T+D_a D_a^T,\nonumber\\
&\Xi_B=\mathcal{H} \otimes PF_b (\mathcal{H} \otimes PF_b)^T+D_bK(D_bK)^{T},\nonumber\\
&\Xi_{1,2}=-\mathcal{H} \otimes P\bar{B}K-\mathcal{H} \otimes PF_b-D_bK,\nonumber\\
&\Xi_{2,1}=-\mathcal{H}^T \otimes K^T\bar{B}^TP-\mathcal{H}^T \otimes F_b^T P-K^TD_b^T.\nonumber
\end{align}
\emph{Proof: }By following the analytical derivation steps used to prove Theorem 1, we conclude that the tracking error system in (\ref{10}) has robust stability if
\begin{align}
\hspace{0cm}\bar\Psi+\Delta \Psi<0,\nonumber
\end{align}
where
\begin{align}
&\Delta \Psi=\begin{bmatrix} Tr(I_N\otimes P\Delta A-\mathcal{H}\otimes P\Delta BK) &-\mathcal{H} \otimes P\Delta{B}K \\ -\mathcal{H}^T \otimes K^T\Delta{B}^TP & *  \end{bmatrix}.\nonumber
\end{align}
Note that the uncertainties can be reformulated as
\begin{eqnarray}
&&\hspace{-0.9cm}\Delta A=F_aV_a(t)D_a,\Delta B=F_bV_b(t)D_b,\label{18}
\end{eqnarray}
where $V_a(t)$ and $V_b(t)$ are unknown time-varying matrices satisfying $V_a(t)^TV_a(t)\le I$ and $V_b(t)^TV_b(t)\le I$, while $F_a,F_b,D_a\in \mathbb{R} ^{3\times 3}$, and $D_b\in \mathbb{R} ^{3\times 1}$ are constant matrices.

Combining (\ref{18}) and Lemma 1, and choosing $\Lambda =I$, we have
\begin{align}
Tr(I_N\otimes P\Delta A)&=Tr(I_N\otimes PF_aV_a(t)D_a)\le\Xi_A,\nonumber
\end{align}
\begin{align}
Tr(\mathcal{H} \otimes P\Delta BK)&=Tr(\mathcal{H} \otimes PF_bV_b(t)D_bK)\le\Xi_B.\nonumber
\end{align}
Applying the results of the above inequalities, if conditions in (\ref{19}) are satisfied, it yields
\begin{align}
\hspace{0cm}\bar\Psi+\Delta \Psi\le\tilde\Psi<0.\nonumber
\end{align}
Thus, the tracking error system in (\ref{10}) has robust stability, and we can easily obtain the controller gain $K$ with reference to Remark 1, which is omitted here.

\emph{Remark 2: }With the starting point of Theorem 1, the co-design framework of a robust controller and a DETM is constructed in Theorem 2. The value of $\varsigma _{1}$ is not only influenced by $c$ and $P$ but also involves trade-offs with event-triggering mechanism and the degree of parameter uncertainty. Additionally, the values of $D_1$ and $T_1$ are primarily concerned with $\varsigma _{1}$, $\varsigma _{2}$ which corresponds to resilience to DoS attacks.
\subsection{Exclusion of Zeno Phenomenon}
To ensure the regular operation of the event-triggered scheme, we must ensure that the scheme can eliminate Zeno phenomenon.

Using the the dynamic and resilient ETM $(\ref{11})$, for $t_{k}^{i}\in \mathcal{A} ( t_{1}, t_{2} )$, we have $t_{k+1}^{i}-t_{k}^{i}=h>0$. Apparently there is no Zeno phenomenon.

Then, for $t_{k}^{i}\in \mathcal{S} ( t_{1}, t_{2} )$, suppose the Zeno phenomenon occurs at time $t_z$. We have $\lim_{k \to \infty } t_{k}^{i}=t_z$. For any $\gamma>0$, there exists an integer $N_\gamma>0$ such that
\begin{eqnarray}
&&\hspace{-0.9cm}t_{k}^{i}\in  [ t_{z}-\gamma , t_z ),\forall k\ge N_\gamma.\label{17}
\end{eqnarray}

With the triggering law $(\ref{3})$ and trigger condition $(\ref{4})$, for $t\in [t_{k}^{i},t_{k+1}^{i})$, we have
\begin{eqnarray}
&&\hspace{-0.9cm}\left \| \varepsilon _{i}(t)  \right \|^{2}\le (\beta _{2} \left \| q _{i}(t)  \right \|^{2}-\varphi _{i} \theta _{i} (t))/\beta _{1}.\nonumber
\end{eqnarray}
The derivative of the state estimation error $\varepsilon _{i}(t)$ satisfies
\begin{eqnarray}
&&\hspace{-0.98cm} \| \dot{\varepsilon } _i(t)\|  \le  \| \tilde{A}\|  \| \varepsilon _i(t) \| + \|  \tilde{B}K\|  \| q_i(t) \|.\nonumber
\end{eqnarray}
From $(\ref{12})$ and $(\ref{6})$, the internal dynamic variable $\theta _{i} (t)$ and tracking error $e_i(t)$ are bounded. Further we get that both $\| \varepsilon _{i}(t) \|$ and $\| \dot{\varepsilon } _i(t)\|$ are bounded, that is,
\begin{eqnarray}
&&\hspace{-0.9cm} \| \varepsilon _{i}(t) \| \le O_1,\| \dot{\varepsilon } _i(t)\|\le O_2.\nonumber
\end{eqnarray}
Note that a sufficient condition for $\pi _{i}(t) \ge 0$ is
\begin{eqnarray}
&&\hspace{-0.9cm} \| \varepsilon _{i}(t)   \|\ge\sqrt{(\beta _{2} \left \| q _{i}(t)  \right \|^{2}-\varphi _{i} \theta _{i} (t))/\beta _{1}}\nonumber\\
&&\hspace{0.23cm}>  \sqrt{\varphi _i\theta_i(0)/\beta _1} e^{-\frac{1}{2}(\varrho _i+\varphi _i\eta _i)t}.\nonumber
\end{eqnarray}
Let $\gamma =  \frac{\sqrt{\varphi _i\theta_i(0)/\beta _1} e^{-\frac{1}{2}(\varrho _i+\varphi _i\eta _i)t_z}}{2O_2} $. According to the hypothesis, we have that ${t_{k+1}^{i} }^{-} < t_z$, where ${t_{k+1}^{i} }^{-}$ is the left limit of $t_{k+1}^{i}$. Then, it follows that
\begin{eqnarray}
&&\hspace{-0.9cm} \| \varepsilon _i({t_{k+1}^{i} }^{-})   \|> \sqrt{\varphi _i\theta_i(0)/\beta _1} e^{-\frac{1}{2}(\varrho _i+\varphi _i\eta _i)t_z}.\nonumber
\end{eqnarray}
In addition,
\begin{eqnarray}
&&\hspace{-0.9cm} t_{k+1}^{i} -t_{k}^{i}\ge {t_{k+1}^{i} }^{-}-t_{k}^{i}\nonumber\\
&&\hspace{0.58cm}>\frac{\sqrt{\varphi _i\theta_i(0)/\beta _1} e^{-\frac{1}{2}(\varrho _i+\varphi _i\eta _i)t_z}}{O_2} =2\gamma.\nonumber
\end{eqnarray}
This contradicts $(\ref{17})$. Thus, there is no Zeno phenomenon.
\section{Simulation Results}
Considering a platoon with five vehicles, comprising one leading vehicle and four follower vehicles. The communication topology obeys a two predecessor single following (TPSF) structure, which is depicted in Fig. 4.
\begin{figure}[H]
\includegraphics[scale=0.5]{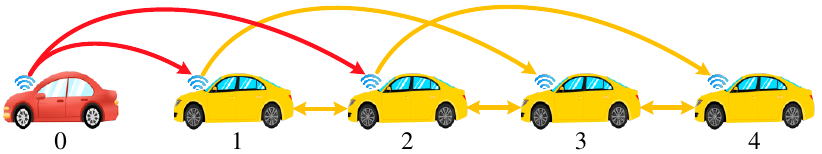}\centering
\caption{TPSF topology of vehicle platoon.}
\end{figure}
The prescribed inter-vehicle spacing is set as $l =10m$. The original states of the leading vehicle and follower vehicles are designed as $x_0=[20,15,0]^T,x_1=[8,16,0]^T,x_2=[-1,14.5,0]^T,x_3=[-9.5,15,0]^T,x_4=[-19,15.5,0]^T$. The leader vehicle is designed to follow an ideal trajectory, encompassing both an acceleration phase and deceleration phase, as illustrated by
\begin{eqnarray}
&&\hspace{-0.9cm}v_0(t)=\left\{\begin{matrix} 15m/s\hfill\hfill & 0s\le t< 10s\\
                                (15+2t)m/s\hfill\hfill &10s\le t< 15s \\
                                 25m/s\hfill\hfill & 15s\le t< 35s\\
                                 (25-2t)m/s\hfill\hfill & 35s\le t< 40s\\
                                  15m/s\hfill\hfill &40s\le t\le 55s\hfill\nonumber\end{matrix}\right.
\end{eqnarray}

Firstly, we validate the efficacy of the proposed method under DoS attacks and parameter uncertainty. Considering that the power-train time constant ${\tau}$ changes due to environmental changes in the driving process, as shown in Fig. 5(a), with the nominal value $\bar{\tau}=0.5s$, and $\Delta \tau$ is raging in $[-0.16s,0.32s]$. From (\ref{20}), we have $\bar{\varpi }=2s^{-1}$ and $\Delta{\varpi }\in [-0.78s^{-1},0.94s^{-1}]$, to determine the constant matrices $F_a=F_b=$ diag$\{0,0,4.7\}$, $D_a=$ diag$\{0,0,-0.2\}$, and $D_b=$ $[0,0,-0.2]^T$. The parameters of dynamic event-triggered mechanism are given as $\varphi _{i}=0.33,\varrho _{i}=0.5,\eta _{i}=0.5,\theta _i=200,h=0.02$. By solving the inequality (\ref{19}), we have $\beta_1=319.38, \beta_2=5.62, K=[4.81,9.12,2.97]$. Further, the parameters related to the tracking error system convergence rate are obtained as $\varsigma _{1}=0.41,\varsigma _{2}=0.27,\varsigma _{*}=0.35,\phi =5.2$. The time of the simulation is set to $[0s,55s]$. The intervals of DoS attacks are represented by shaded areas in the simulation diagram, as shown in Fig. 5(b). By selecting $T_1=2,D_1=6$, and through calculation, we obtain $T_2>9.46$ and $D_2>11.33$. This yields $\left |\mathcal{A} \left ( 0, 55 \right )  \right | =10< 10.85$ and $\mathcal{N} \left ( 0,55 \right )=7<7.81$, satisfying the constrains for DoS attacks.
\begin{figure}[h]
\centering
\subfigure[Power-train time constant]
{
		\includegraphics[scale=0.49]{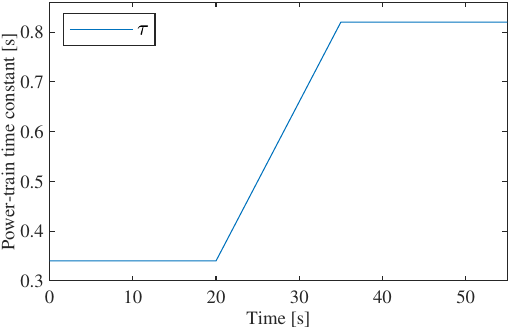}}
\subfigure[Attacked intervals]
{
		\includegraphics[scale=0.49]{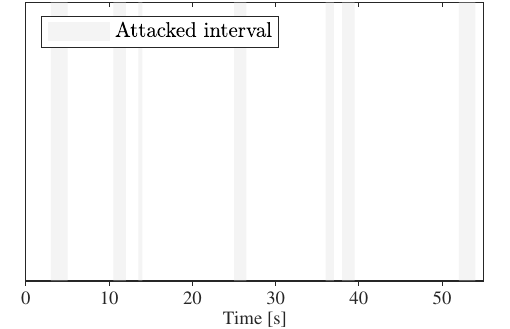}}
\caption{Settings of parameter uncertainty and DoS attacks.}
\end{figure}

With the above parameters, the position, relative positions to the leading vehicle, velocity, and evolution of internal dynamic variable of each vehicle are illustrated in Fig. 6. Specifically, Fig. 6(a) demonstrates the maintenance of a cohesive platoon during the simulation. Additionally, Fig. 6(b) and Fig. 6(c) indicate that the following vehicles can track the speed changes and maintain the prescribed distance of the leading vehicle with DoS attacks and parameter uncertainty. Fig. 6(d) indicates that the internal dynamic variable can be adjusted based on the adjacent vehicle states and the self-state. The triggering instants of the following vehicles are presented in Fig. 7, which shows that the proposed resilient and dynamic event-triggered mechanism can save communication resources and exclude Zeno phenomenon.
\begin{figure}[h]
\centering
\subfigure[Positions]
{
		\includegraphics[scale=0.48]{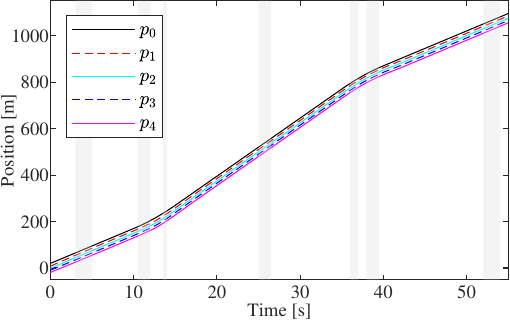}}
\subfigure[Relative positions]
{
		\includegraphics[scale=0.48]{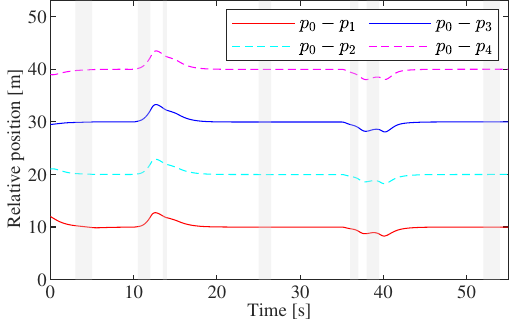}}
\\
\subfigure[Velocities]
{
		\includegraphics[scale=0.48]{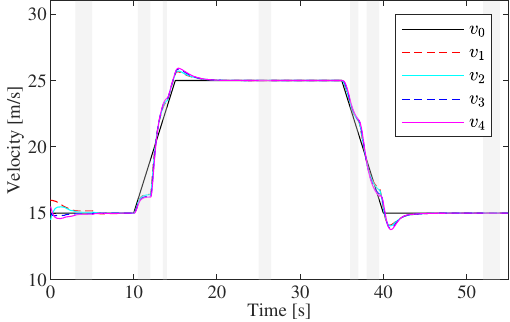}}
\subfigure[Internal dynamic variables]
{
		\includegraphics[scale=0.48]{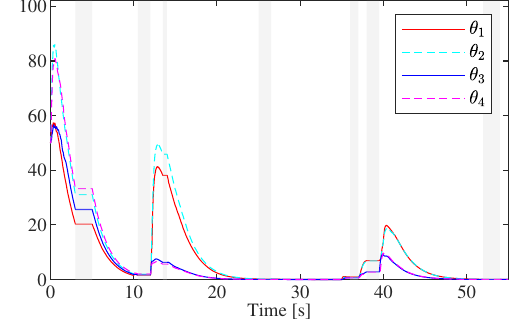}}
\caption{Platoon behavior under DoS attacks and parameter uncertainty.}
\end{figure}

\begin{figure}[h]
\centering
{
		\includegraphics[scale=0.47]{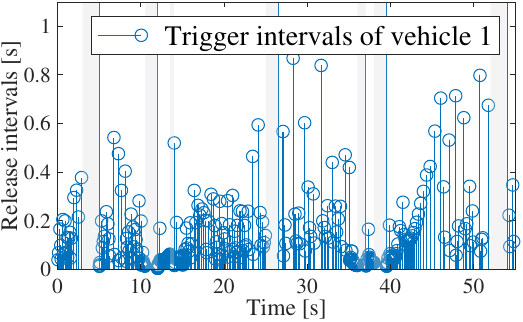}}
{
		\includegraphics[scale=0.47]{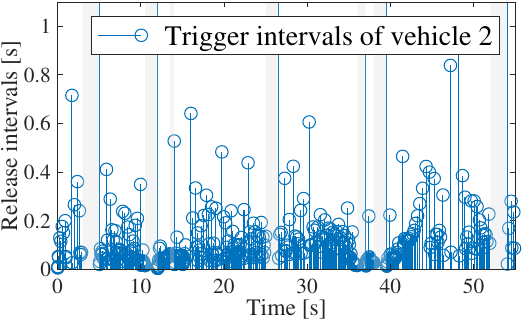}}
\\
{
		\includegraphics[scale=0.47]{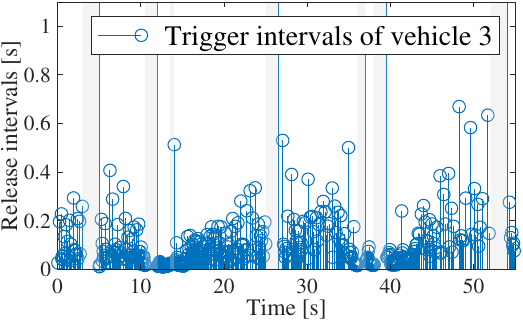}}
{
		\includegraphics[scale=0.47]{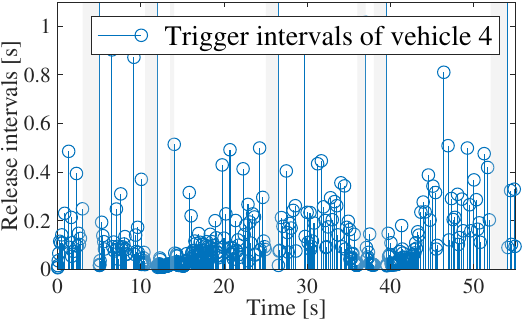}}
\caption{Release instants of each following vehicle.}
\end{figure}

To show the superiority of our proposed method in saving communication resources, we present the average tracking error and count the number of triggering times for different event-triggered approaches in the absence of DoS attacks and parameter uncertainty, as shown in Fig. 8 and Table I. The results indicate that the proposed DETM in (\ref{4}) sacrifices a little control performance to have much fewer triggering times than those in (\ref{21}) and \cite{xiao2022resource}, facilitated by the use of the internal dynamic variable.
\begin{figure}[h]
\centering
{
		\includegraphics[scale=0.46]{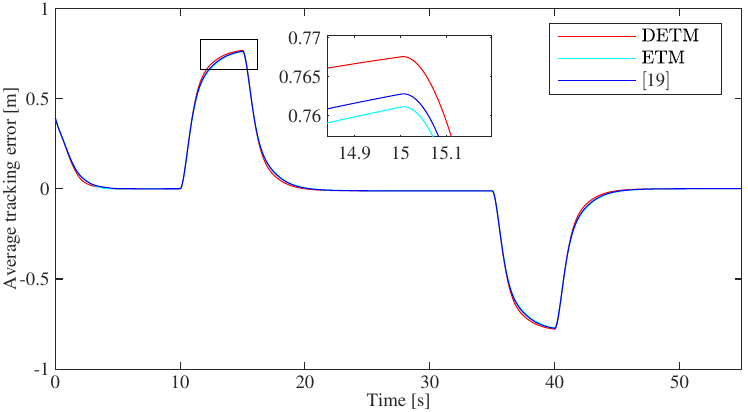}}
\caption{Average tracking errors under various triggering mechanisms.}
\end{figure}
\begin{table}[H]
    \renewcommand\arraystretch{1.2}
    \caption{Comparisons of triggering times}\label{table1}
    \resizebox{\linewidth}{!}{
    \centering
    \begin{tabular}{c c c c c}
        \toprule
        Event-triggered mechanisms & vehicle 1 & vehicle 2 & vehicle 3 & vehicle 4 \\
        \midrule
        DETM in (6) & 331 & 352 & 419 & 397 \\
        \midrule
        ETM in (8) & 577 & 426 & 601 & 628 \\
        \midrule
        ETM in \cite{xiao2022resource} & 493 & 381 & 522 & 495 \\
        \bottomrule
    \end{tabular}}
\end{table}
To demonstrate the effectiveness of our proposed method in ensuring robustness under parameter uncertainty, we design the controllers in two cases: $\tau$ remains constant at $0.5s$, and $\tau$ varies within the interval $[0.34s, 0.82s]$. When $\tau$ remains constant, the method proposed in Theorem 1 is used for controller design. Conversely, when $\tau$ varies, the methods described in Theorem 2 and in \cite{li2017robustness} are employed. Based on the conditions in Fig. 5(a), we get the average tracking error as depicted in Fig. 9. During $[0s,20s]$, the value of $\tau$ is $0.34s$, and the control performance of the method proposed in Theorem 2 is similar to that of the method in \cite{li2017robustness}. During $[35s,55s]$, the value of $\tau$ is $0.82s$, and the method proposed in Theorem 2 shows stronger robustness under parameter uncertainty than the method in \cite{li2017robustness}. The fact is that the proposed method is less conservative than that in \cite{li2017robustness}, where the controller is designed according to the lower bound of the parameter uncertainty. In contrast, the controller designed without considering the existence of parameter uncertainty shows poor robustness.

\begin{figure}[h]
\centering
{
		\includegraphics[scale=0.59]{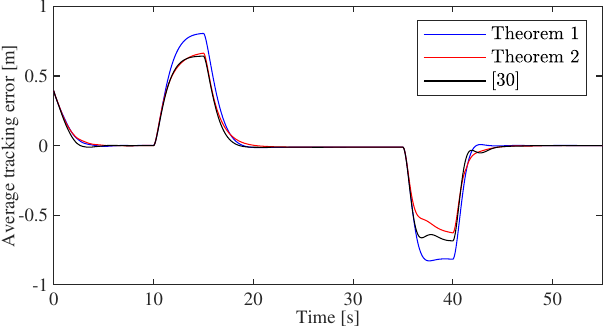}}
\caption{Average tracking errors using different controllers.}
\end{figure}

Parameter $\varsigma_{1}$ is related to the rate of convergence. To find a clear understanding of the impact of DoS attacks, bounded parameter uncertainty, and DETM parameters on platoon system control performance, initially we neglect the parameter uncertainty and the impact of DoS attacks to get the values of $\varsigma_{1}$ in  predecessor following (PF) topology, TPSF topology, and bidirectional predecessor following (BPF) topology  with different numbers of platoon vehicles using the design method in Theorem 2, represented by $PF_n,TPSF_n,BPF_n$. Moreover, we keep the related triggering parameters unchanged and only consider the influence of parameter uncertainty, and the obtained value of $\varsigma_{1}$ is represented by $PF_u,TPSF_u,BPF_u$. Similarly, we set more conservative triggering parameters to get the value of $\varsigma_{1}$, denoted as $PF_c,TPSF_c,BPF_c$. We also get the $\varsigma_{1}$ in the presence of DoS attacks, represented by $PF_a,TPSF_a,BPF_a$.
\begin{figure}[h]
{
    \begin{center}
        \includegraphics[scale=0.39]{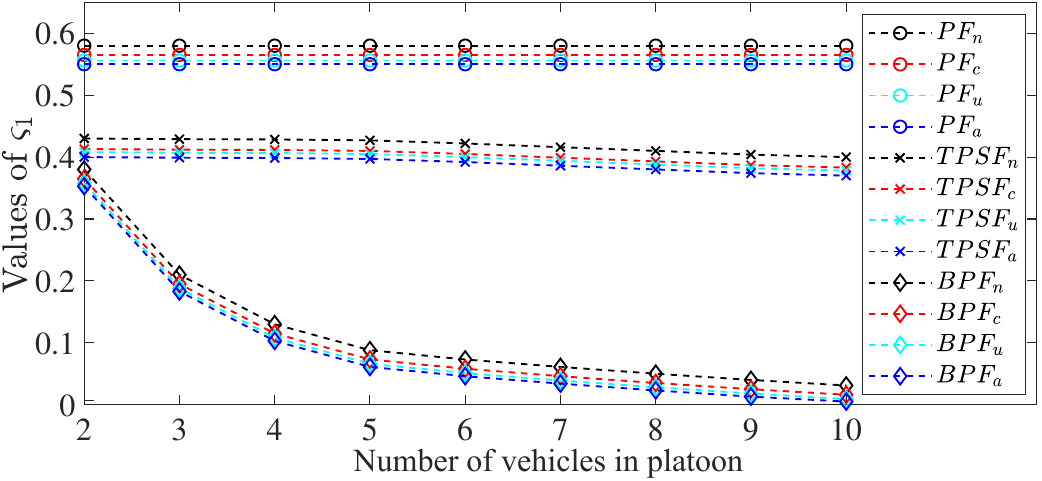}
    \end{center}
}
\caption{Comparison of $\varsigma_{1}$ under different conditions.}
\end{figure}

Considering different communication topologies, Fig. 10 gives a comparison of $\varsigma_{1}$ under different conditions, revealing that the parameter uncertainty, the DoS attacks, and the adjustment of triggering parameters will affect the control performance. Particularly, results demonstrate the performance superiority of the proposed method due to collaborative design and analysis. Additionally, different communication topologies also have an impact on platoon control performance.
\section{Conclusion}
This work focuses on the application of DTEM for vehicle platoon control systems under DoS attacks and parameter uncertainty. Initially, the value of power-train time instant caused by various driving conditions is modeled as parameter uncertainty. Then, considering the influence of DoS attack and limited network bandwidth, a resilient and dynamic ETM is designed. Subsequently, a co-design framework of a robust controller and a DETM is constructed, in which the resilience of the platoon system to DoS attacks is analyzed under the premise that robustness is ensured, with the analysis procedure eliminating Zeno behavior. Finally, extensive simulation results show that the design method effectively maintains the performance of control in the presence of DoS attacks and parameter uncertainty while also saving communication resources. The future work include addressing the control challenges of heterogeneous platoons under compound cyber attacks, enhancing the robustness of mixed platoon control, and tackling platoon control problems in special scenarios such as mines, seaports, and airports.

\end{spacing}
\begin{spacing}{1.4}

\end{spacing}
\end{document}